# Fluorine-based color centers in diamond

S. Ditalia Tchernij[1,2,3], T. Lühmann[4], E. Corte[1] , F. Sardi[1], F. Picollo[1,2], P. Traina[3], M. Brajkovic[5], A. Crnjac[5], S. Pezzagna[4], I. P. Degiovanni[3,2], E. Moreva[3], P. Aprà[1,2], P. Olivero[1,2,3], Z. Siketić[5], J. Meijer[4], M. Genovese[3,2], J. Forneris[1,2,4,*]

[1] Physics Department, University of Torino, Torino 10125, Italy
[2] Istituto Nazionale di Fisica Nucleare (INFN), sezione di Torino, Torino 10125, Italy
[3] Istituto Nazionale di Ricerca Metrologica (INRiM), Torino 10135, Italy
[4] Department of Nuclear Solid State Physics, Universität Leipzig, Leipzig 04109, Germany
[5] Laboratory for Ion Beam Interactions, Ruđer Bošković Institute, Zagreb 10000, Croatia

**Abstract**
We report on the creation and characterization of the luminescence properties of high-purity diamond substrates upon F ion implantation and subsequent thermal annealing. Their room-temperature photoluminescence emission consists of a weak emission line at 558 nm and of intense bands in the 600 – 750 nm spectral range. Characterization at liquid He temperature reveals the presence of a structured set of lines in the 600 – 670 nm spectral range. We discuss the dependence of the emission properties of F-related optical centers on different experimental parameters such as the operating temperature and the excitation wavelength. The correlation of the emission intensity with F implantation fluence, and the exclusive observation of the afore-mentioned spectral features in F-implanted and annealed samples provides a strong indication that the observed emission features are related to a stable F-containing defective complex in the diamond lattice.

* Corresponding author: jacopo.forneris@unito.it

## 1. Introduction

Diamond has been widely investigated as an appealing material for quantum optics and quantum information processing applications, due to the availability of several classes of optically active defects (usually referred to as "color centers") that can be suitably engineered in its crystal structure [1-3]. To date, the most prominent type of defect is the so-called negatively-charged nitrogen-vacancy center (NV$^-$), due to several key features of this system, namely: photo-stability at room temperature, high quantum efficiency and most importantly unique spin properties with great potential for applications in quantum sensing and computing [4-9].

The need for single-photon emitters displaying desirable opto-physical properties (high emission rate, narrow linewidth) has also motivated the discovery and characterization of several classes of optical centers in diamond alternative to the NV complex, based (among others) on group-IV impurities [10] (Si [11,12], Ge [13,14], Sn [15,16], Pb [17,18]) and noble gases (He, Xe [19-21]). In this context, ion beam implantation represents a powerful and versatile tool to engineer a broad range of different types of color centers, allowing for the fine control of key parameters such as ion species and energy, as well as irradiation fluence to determine the type and density of defect complexes [22, 23].

Up to now though, the number of the emitters characterized by a reproducible fabrication process is fairly limited, and a systematic investigation in this field is still to be finalized. Thus, the fabrication of novel luminescent defects with desirable properties upon the implantation of selected ion species still represents a crucial strategy to achieve further advances in the fields. Following from a previous report with preliminary results on this system [24], in this work we report on the systematic characterization in photoluminescence (PL) under different optical excitation wavelengths of F-related color centers in diamond created upon ion implantation and subsequent annealing.

## 2. Experimental

The experiments were performed on a type-IIa single-crystal diamond sample produced by ElementSix with Chemical Vapor Deposition technique, namely a $2\times2\times0.5$ mm$^3$ "electronic grade" substrate having nominal concentrations of both substitutional nitrogen and boron impurities below the 5 ppb level.

The sample was implanted with 50 keV F$^-$ ions at the low-energy accelerator of the University of Leipzig. Several circular regions of ~175 μm diameter were irradiated at varying fluences in the $5\times10^{10} - 5\times10^{15}$ cm$^{-2}$ range by the use of a custom beam collimator. An additional region was implanted with 1.47 MeV F$^{2+}$ ions ($1\times10^{13}$ cm$^{-2}$ fluence) at the Laboratory for Ion Beam Interactions of the Ruđer Bošković Institute. The sample was then processed with a high-temperature thermal annealing (1200 °C, 4 hours in vacuum at ~10$^{-6}$ mbar pressure) and a subsequent oxygen plasma treatment, with the purpose of minimizing the background fluorescence from surface contaminations.

A preliminary PL characterization was performed using a Horiba Jobin-Yvon HR800 micro-Raman spectrometer equipped with a CW excitation laser at 532 nm wavelength.

Temperature-dependent PL measurements were performed using a custom-made cryogenic single-photon confocal microscope exploiting a closed-cycle optical cryostat operating in the 4 – 300 K temperature range. Optical excitation was provided through a 100× air objective (0.85 NA) by a set of CW laser diodes with emission wavelengths of 405 nm, 488 nm and 520 nm. A suitable set of optical filters enabled the detection of PL emission at wavelengths larger than 550 nm. The PL spectra were acquired using a single-grating monochromator (1200 grooves mm$^{-1}$, 600 nm blaze) fiber-coupled to a Si-single-photon-avalanche photo-diode operating in Geiger mode.

## 3. Results

### 3.1 PL emission at room temperature

Ensemble PL spectra acquired at room temperature are presented in Fig 1 for different F implantation fluences.

The emission spectra reported in Fig. 1a were acquired from the implanted regions under 532 nm laser excitation (26.2 mW laser power) and exhibit a broad, intense and apparently unstructured PL band in the 600 – 800 nm spectral region (fluorine-related band, FB1, in the following), with a maximum centered at ~670 nm. Although not particularly characteristic in its spectral features, the observation of this band is consistent with what reported in Ref. [24] under 488 nm excitation, where a broad emission band in the 590-800 was formed upon F implantation at a 5x10^14 cm^-2 ion fluence and a subsequent thermal annealing at 1600 °C.

It is worth mentioning that the FB1 band is partially spectrally overlapping with the emission of the NV^0 and NV^- center in diamond (i.e. PL bands in the 575-650, and 638 – 750 nm range, respectively, typically observed together in ensemble PL spectra [4]). To rule out the possibility that the observed spectral feature can be attributed to NV-related emission, a pristine region of the same sample was irradiated with C^- ions at 35 keV energy (5×10^15 cm^-2 fluence) and processed alongside the F-implanted region with the same annealing parameters. The PL spectrum from this test area, acquired under the same excitation conditions, is reported in Fig. 1b for the sake of comparison. It displays a significantly weaker band centered at 640 nm, whose intensity is comparable with the second-order Raman scattering of diamond observed at ~620 nm (corresponding to a 2664 cm^-1 Raman shift), while not exhibiting any features similar to those observed for the FB1 band. This observation is indicative of the fact that the FB1 band is neither related to the formation of NV centers upon the introduction of lattice vacancies in a N-containing diamond substrate, nor to generic ion-induced structural damage.

In addition, the substantial increase (Fig. 1a) of the FB1 intensity at increasing F implantation fluences supports our attribution of this spectral feature to a stable F-containing defect in the diamond lattice.

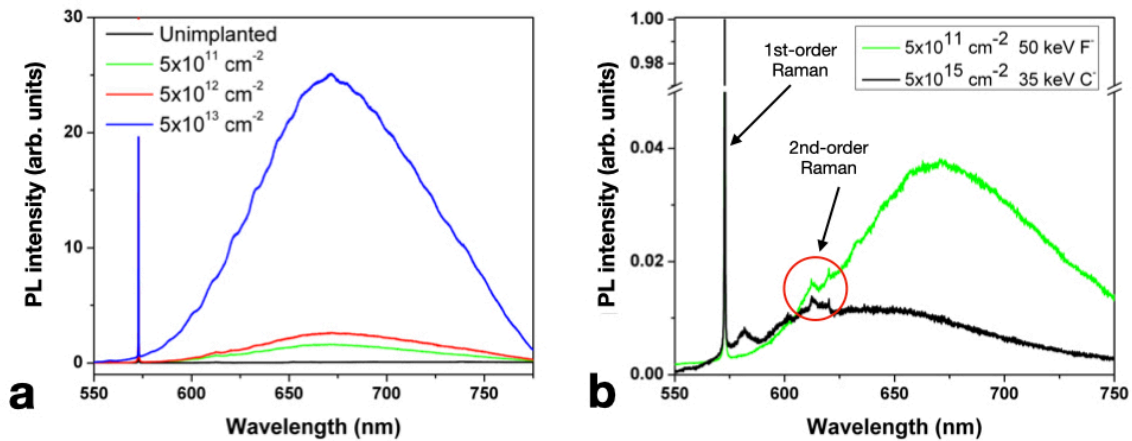

**Fig 1. a)** PL spectra acquired under 532 nm laser excitation wavelength from regions implanted with different 50 keV F- ion fluences in the 5×10^11 - 1×10^13 cm^-2 range. A PL spectrum of pristine spectrum is included for the sake of comparison (black line). b) Comparison of the PL spectrum acquired from the region implanted with 5×10^11 cm^-2 50 keV F- ion fluence (green line) with a sample region irradiated with 35 keV C^- ions at 5×10^15cm^-2 fluence (black line). Both curves are normalized to the first-order Raman peak. The first-order and second-order Raman features of diamond are highlighted.

Fig. 2a displays a zoom-in in the 540-640 nm spectral range of the PL data reported in Fig. 1a. A weak PL peak at 558 nm appears in the regions implanted with F ions at fluences higher than 5×10^12 cm^-2. This peak was consistently observed in all of the room-temperature measurements performed on the F-implanted regions, while being absent from the test spectrum reported in Fig. 1b. While the peak position might be compatible with a 557 nm feature observed in cathodoluminescence from electron-irradiated diamonds that was previously attributed to an interstitial-type defect (ITD in the following) [19, 25], a different interpretation is suggested by the persistence of the peak upon 1200 °C annealing (ITD is reported to anneal out at 700 °C [26]). Furthermore, it is worth remarking that this spectral feature was not reported in a previous PL characterization of C-implanted diamond [27], and analogously has not been observed in our measurements in nominally identical samples implanted with a wide range of different ion species. Based on these considerations, a tentative attribution of the 558 nm line as related to an F-containing defect is made, although its relationship with the FB1 band is unclear on the basis of the available dataset.

Finally, Fig. 2b shows a PL spectrum acquired under 520 nm laser excitation (5.6 mW power, 565 nm long-pass filter) from the region implanted with 1.47 MeV $F^{2+}$ ions at $1 \times 10^{13}$ cm$^{-2}$ fluence. Since the spectrum was acquired with the cryogenic single-photon confocal microscope rather than with the micro-Raman spectrometer, the PL intensity scale cannot be quantitatively compared with that of the previously discussed measurements. However, it is worth noting that Fig. 2b displays the same spectral features reported in Fig. 1a. This observation further corroborates the attribution of the emission spectrum to a F-containing lattice defect, and indicates that such optical center can be fabricated upon the implantation of fluorine ions of both keV and MeV energies (possibly with different creation yields, whose investigation is beyond the scope of this work), thus ruling out its possible attribution to intrinsic lattice defects related to high-energy irradiation effects.

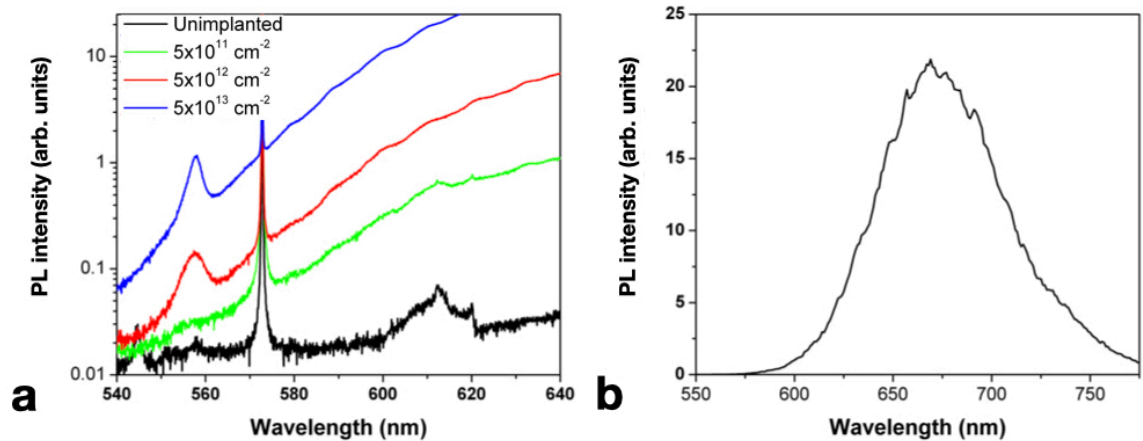

**Fig 2. a)** Inset of Fig. 1a in the 540-660 nm spectral range. The axes are rescaled to highlight the PL emission peak at 558 nm. **b)** PL spectrum acquired under 520 nm laser excitation from a sample region implanted with 1.47 MeV $F^{2+}$ ions at $1 \times 10^{13}$ cm$^{-2}$ fluence.

### 3.2 PL emission at variable temperatures

In order to gain a deeper insight into the spectral features of F-related color centers with respect to the aforementioned results and the previous report in Ref. [24], low-temperature PL measurements were performed at different excitation wavelengths. Fig. 3 displays an ensemble PL spectrum acquired at 4.5 K under 488 nm excitation (3.6 mW laser power) from a region implanted at $5 \times 10^{13}$ cm$^{-2}$ fluence. At the reported temperature, the FB1 band reveals an articulated structure of different emission lines, the most prominent being located at 600, 611, 623, 634, 647, 658 and 671 nm. In Fig. 4, PL spectra acquired from the same region both at 4.5 K and at room temperature are reported for different laser excitation wavelengths, including 405 nm (10 mW laser power) and 520 nm (2.4 mW). It is worth noting that the spectra acquired using the 405 nm laser were limited to the 500-700 nm spectral range due to the luminescence of optical elements employed in the confocal microscopy setup at higher wavelengths. The persistence of the emission lines at the same wavelengths under different excitation sources indicates that they cannot be attributed to Raman scattering. Furthermore, they can hardly be attributed to any interference process due to multiple internal reflections within the irradiated sample, firstly because they are visibly temperature-dependent, and most importantly because (differently from what is reported in Ref. [28]) the implantation fluence is low enough to avoid the presence of a reflective layer below the sample surface. Within a ~3 nm limit due to the spectral resolution of the experimental apparatus, all of the aforementioned lines are uniformly spaced in energy of $\Delta E = (36 \pm 5)$ meV. If the 600 nm emission line corresponds, as assumed here, to the zero-phonon line (ZPL) of the F-related center, such energy shift value is compatible with an electron-phonon coupling, as it falls in the energy range (10-100 meV) related to quasi-local vibrations [19].

Remarkably, the 558 nm line reported in Fig. 2a is also visible in Fig. 3, although its relative intensity with respect to the FB1 band maximum is further decreased with respect to room-temperature conditions. Such observation is indicative of the fact that the 558 nm emission can hardly be attributed to the ZPL of the FB1 band. This interpretation is further supported by the 155 meV energy separation with respect to its closest emission line at 600 nm. Therefore, the available experimental data are not sufficient to justify any reasonable attribution of the observed peak.

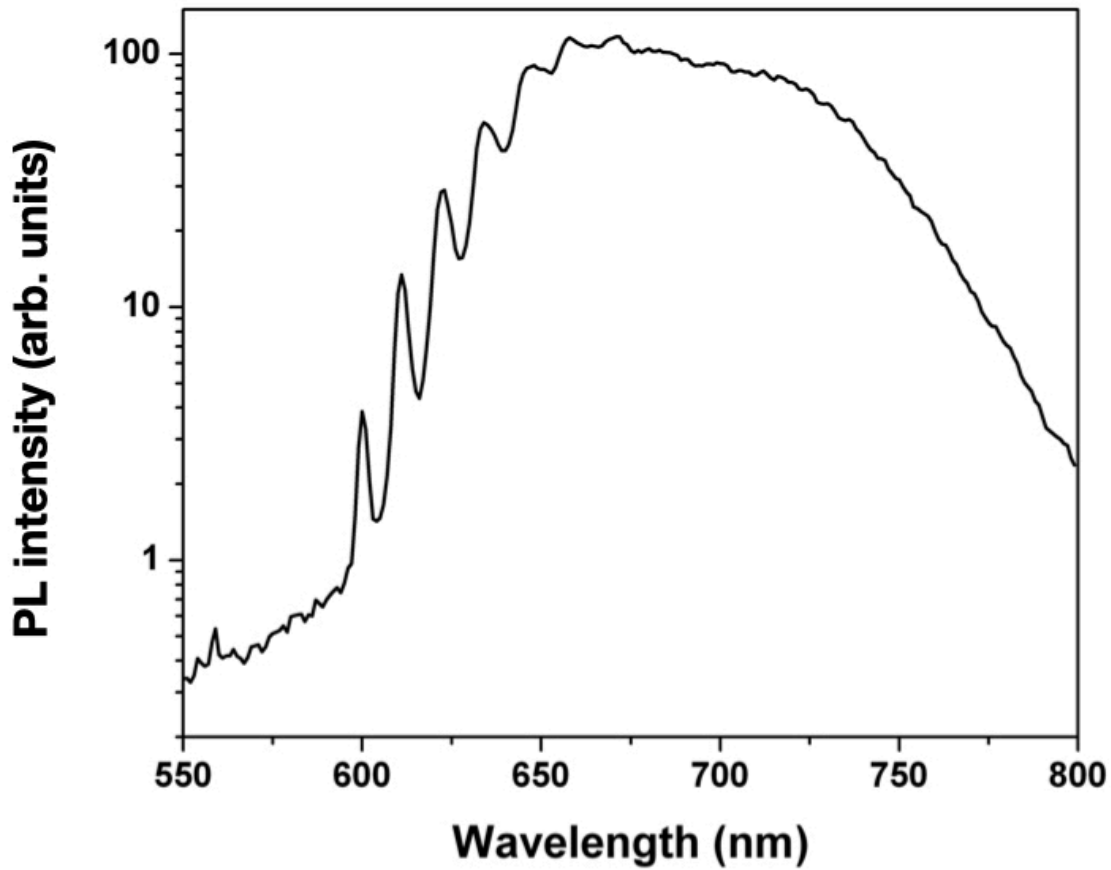

**Fig 3.** PL spectrum acquired at 4.5 K temperature under 488 nm laser excitation from the region of sample #1 implanted with at $5\times10^{13}$ cm$^{-2}$ fluence.

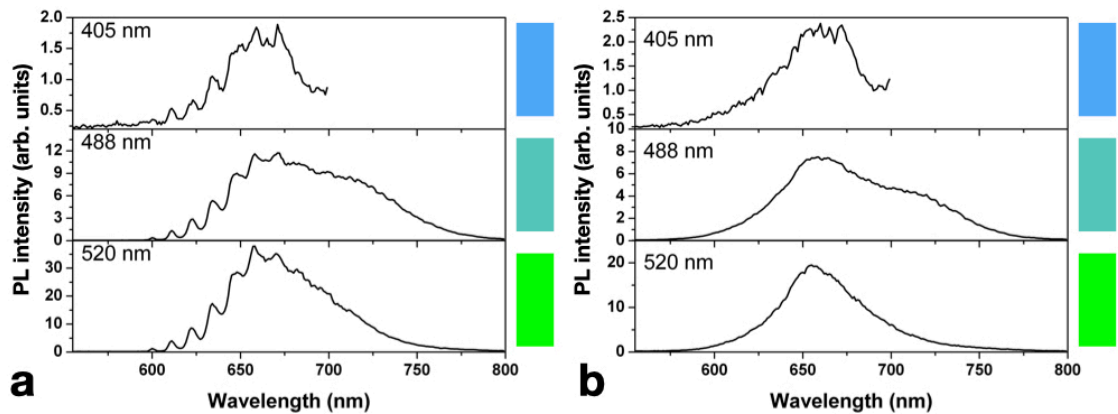

**Fig 4.** Ensemble PL spectra acquired at 4.5 K **(a)** and room temperature **(b)** from the region of sample #1 implanted at a fluence of $5\times10^{13}$ cm$^{-2}$ under different excitation wavelengths.

While no significant differences were observed between the PL peaks reported in Fig. 4, a comparison between the spectra reveals an additional emission band (FB2 in the following), which is visible (both at 4.5 K and 300 K) under 488 nm excitation only. This component appears as a second band centered at ~710 nm, not observed in the previous report on F-implanted diamond [24]. The emergence of the FB2 band cannot be determined under 405 nm excitation as it lies outside of the available spectral range. If confirmed, its observation could lead to the attribution of FB1 and FB2 as different charge states of the same defect, in a similar fashion to the well-established interpretation of the emission bands of the neutral and negative charge states of the nitrogen-vacancy complex. In this case, FB2 would be related to a higher excitation energy (> 2.39 eV) with respect to FB1 despite resulting a lower photon energy. This might indicate the presence of indirect transitions or non-radiative processes involved in the luminescence mechanism, to be further assessed at the single-photon emitter level.

Fig. 5 shows the PL spectra at various temperatures in the 4.5 – 300 K temperature range, under 488 nm excitation. The assessment of a trend in the relative intensity of the FB1 and FB2 bands was not possible due to slightly different focussing conditions. Nevertheless, the progressive broadening of the set of lines in the 600 – 670 nm range is clearly visible up to temperatures as high as 160 K, thus providing a further confirmation of their attribution as phonon sidebands of the 600 nm ZPL emission. Conversely, the absence of any distinguishable internal structure of the FB2 band even at 4.5 K temperature cannot be understood without a specific interpretative model, and thus will require further investigations.

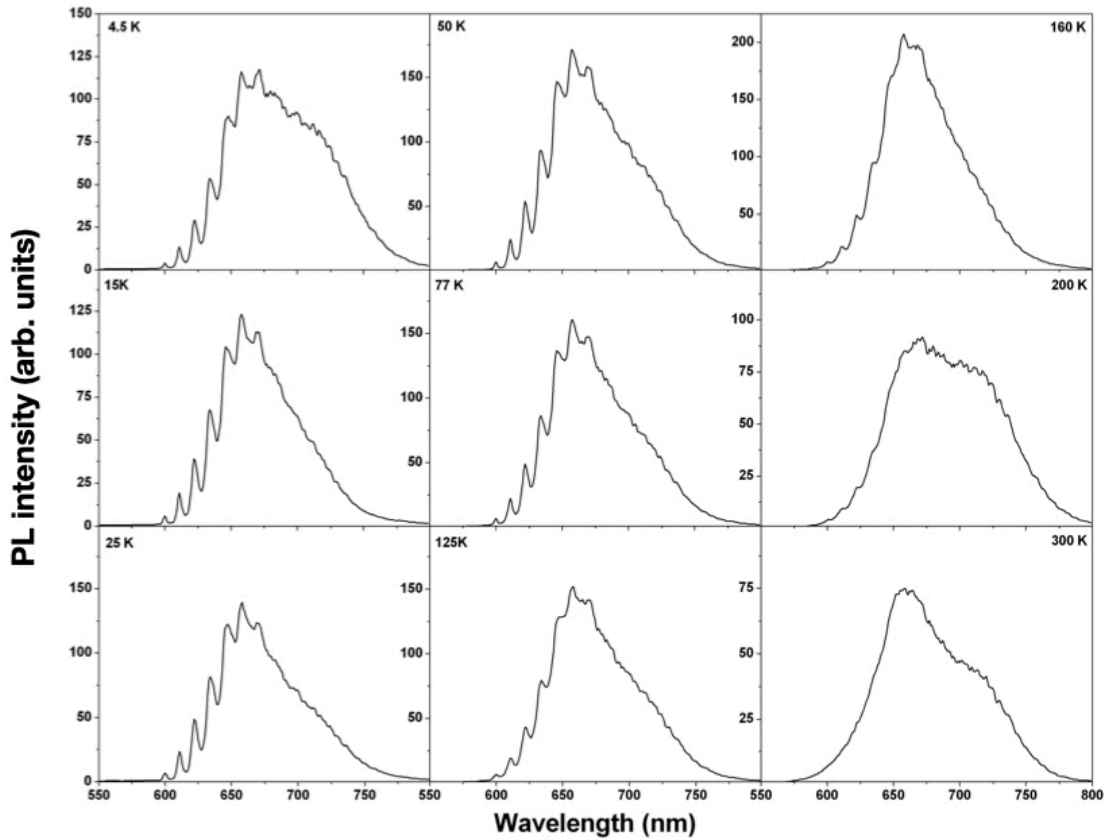

**Fig 5** Ensemble PL emission from the region of sample #1 implanted at a fluence of $5 \times 10^{13}$ cm$^{-2}$ under 488 nm excitation as a function of temperature.

## 3. Conclusions

In this work we reported on the fabrication by means of ion beam implantation and thermal annealing of a new class of color center based on F impurities. The ensemble PL characterization showed several peculiar spectral features that, to the best of the authors' knowledge, were not reported so far at the state of the art, namely: a weak emission peak at 558 nm, two bands centered at ~670 nm (FB1) and ~710 nm (FB2), the latter being visible under 488 nm excitation but not under 520 nm. At cryogenic temperatures, the FB1 band exhibited an articulated structure of separate emission lines, whose arrangement is compatible with the attribution of the 600 nm spectral features to the ZPL emission of the center, coupled to a set of six phonon replica spaced by a $(36 \pm 5)$ meV vibrational energy attributed to quasi-local vibration. The weak 588 nm emission could not be related to the FB1, nonetheless it is not implausible that it originates from a different F-containing lattice defect.

The results discussed in this work offer the first extensive characterization of F-related defects, thus confirming the claim on their optical activity presented in Ref. [24]. In such report, their characterization at the single-photon emitter level was made challenging by the strong similarity in both the spectral signatures and the excited state radiative lifetime with respect to those of the NV center [4]. A systematic investigation on individual centers will also be necessary in order to clarify the attribution of the 558 nm emission line and the nature of the FB2 band.

While the wide emission band of F-implanted diamond is not suitable for the development of monochromatic sources of non-classical light, a full understanding of the centers' properties might open the path towards novel applications in quantum sensing and quantum information processing. Indeed, preliminary *ab initio* simulations, while not aimed at the identification of the defect optical transitions, suggest a ground state electronic spin configuration similar to that of the NV center (S=1/2, and S=1 for neutrally- and negatively-charged impurity-vacancy defects). If confirmed by EPR experiments, such property could pave the way to the utilization of F-related centers in diamond for quantum information processing applications, taking advantage of the non-zero nuclear spin of the naturally occurring $^{19}$F for hyperfine interactions for defects control and coupling [29].

## Acknowledgements

This work was supported by the following projects: Coordinated Research Project "F11020" of the International Atomic Energy Agency (IAEA); Project "Piemonte Quantum Enabling Technologies" (PiQuET), funded by the Piemonte Region within the "Infra-P" scheme (POR-FESR 2014-2020 program of the European Union)"; and "Departments of Excellence" (L. 232/2016), funded by the Italian Ministry of Education, University and Research (MIUR); T.L., S.P and J.M. acknowledge the support of the ASTERIQS program of the European Commission. The work was supported by projects EMPIR 17FUN06 "SIQUST" and 17FUN01 "BeCOMe; these projects received funding from the EMPIR Programme cofinanced by the Participating States and from the European Union Horizon 2020 Research and Innovation Programme. S.D.T. and J.F. gratefully acknowledge the EU RADIATE Project (proposal 19001744) for granting transational access to the Laboratories of the Ruđer Bošković Institute.